\documentclass[useAMS,usenatbib,onecolumn]{mn2e}
\usepackage{graphicx}
\usepackage{amsmath}
\usepackage{longtable,array}
\usepackage{lscape}
\usepackage{color}
\usepackage{epsfig}
\usepackage{url}
\usepackage{hyperref}
\usepackage{mwe}
\usepackage{epstopdf}

\newcommand{\cm}{cm$^{-1}$}

\newcommand{\gtot}{g$_{tot}$}

\def\a0{{$a_{\rm 0}$}}

\title[ExoMol line lists  XXXV: Ammonia]{ExoMol molecular line lists XXXV:  a rotation-vibration line list for hot ammonia}
\author[Coles et al]{\large
{Phillip A. Coles, Sergei N. Yurchenko and Jonathan Tennyson\thanks{Email: j.tennyson@ucl.ac.uk}}\\
Department of Physics and Astronomy, University College London, London WC1E 6BT, UK}

\date{\today}

\date{Accepted XXXX. Received XXXX; in original form XXXX}

\pagerange{\pageref{firstpage}--\pageref{lastpage}} \pubyear{2019}

\begin{document}

\maketitle

\label{firstpage}

\begin{abstract}

A new hot line list for $^{14}$NH$_3$ is presented. The line list CoYuTe was
constructed using an accurate, empirically refined potential energy surface and
a CCSD(T)/aug-cc-pVQZ  \textit{ab initio} dipole moment surface of ammonia,
previously reported. The line list is an improvement of
the ammonia line list BYTe [Yurchenko et al., \textit{Mon. Not. R. Astron.
Soc.}, 413, 1828 (2011)].  The CoYuTe line list covers wavenumbers up to 20~000
\cm, i.e. wavelengths beyond 0.5 $\mu$m for temperatures up to 1500~K.
Comparisons  with the high temperature experimental data from the literature
show excellent agrement for wavenumbers below 6000~cm$^{-1}$.  The CoYuTe line
list contains 16.9 billion transitions and is available from the ExoMol website
(\url{www.exomol.com}) and the CDS database.

\end{abstract}

\section{Introduction}

Ammonia is the major nitrogen-containing molecule observable in a
number of astrophysical environments. For example, its spectral
signature has long been observed in the atmospheres of Jupiter,
Saturn, and Titan \citep{77WoTrOw.NH3}. Emissions from hot ammonia
were observed following the collision of comet Shoemaker-Levy 9 with
Jupiter \citep{jt154s}
and recent analysis has shown ammonia absorption features in the
visible spectrum of Jupiter \citep{18IrBoBr.NH3}.  These features were
poorly represented by data available in standard databases but could
be modeled using a preliminary version of the line list presented here
\citep{jt745}.

Although attempts to detect ammonia in the atmosphere of an exoplanet
have so far proved inconclusive \citep{jt495}, it is thought to be an
important component of the chemistry of hot Jupiter exoplanets
\citep{17MaMaxx.NH3}.  On Earth atmospheric ammonia is often
associated with human activity such as biomass burning
\citep{88HeRaHo.NH3}.  It has been proposed as a promosing
biosignature  in H$_2$-dominated atmospheres on rocky exoplanets
\citep{13SeBaHi.NH3}. Line lists, such as the one presented here, are
important for modelling spectra that might be observed in future space
missions \citep{18DaBaLa,jt717s}.

Ammonia has recently been observed in a planet-forming disk
\citep{16SaHoBe.NH3}; circumstellar ammonia spectra can only be
understood using a non-LTE multilevel radiative transfer model which
includes the effects of near-infrared (NIR) radiative pumping through
vibrational transitions \citep{16ScHeSz.NH3}. Similarly, interstellar
ammonia has long been observed to mase \citep{86MaIrMa.NH3} and
non-LTE spectra of ammonia have been observed in comets where its
signature has been seen in fluorescence \citep{13ViMaMu.NH3}.

Detailed ammonia spectra have been observed, and partially assigned, in the
spectra of cool brown dwarfs \citep{11BoBuSi,jt596}; its presence has
been used as a tracer of chemical equilibrium  in T dwarfs \citep{06SaMaCu.NH3}. The spectrum of ammonia is
generally assumed to be signature of  Y dwarfs \citep{12SaMaAb.NH3,jt484}.
However, recent observations of Y dwarfs suggest that the ammonia abundance may
be lower than anticipated  \citep{15LeMoCa.NH3,18MoSkAl}.

Modelling or interpretting the spectrum of hot or non-LTE ammonia requires a substantial quantity of laboratory data. To this end a number studies of ammonia have been performed with the view of producing extensive line lists of spectroscopic transitions
\citep{jt466,jt500,11HuScLea.NH3,11HuScLeb.NH3,12HaLiBe.NH3,15Yurche.NH3,19CoOwKu.NH3}.
Notable amongst these are the BYTe line list of \citet{jt500} and the HSL-pre3  line list \citep{13HuLexx.NH3} of \citet{12SuBrHuSc}.  The
transition frequencies predicted by HSL-pre3 are generally more accurate than those of BYTe but HSL-pre3 does not provide transition
intensities. BYTe was the first, comprehensive line list capable of modelling the opacity and spectrum of hot ammonia. BYTe contains 1.1
billion transitions but becomes increasing less accurate in the near infrared and is largely not useful at visible wavelengths. A preliminary,
low-temperature
version of the CoYuTe line list, called C2018, has been used to successfully model visible absorption by ammonia in Jupiter
\citep{jt745}.

The present paper presents a new line list for hot ammonia called CoYuTe. CoYuTe is constructed as part of the ExoMol project \citep{jt528}
which aims to provide comprehensive line lists for studies of exoplanets and other hot or non-LTE atmospheres. CoYuTe can be seen as the
logical successor to BYTe. It improves on the accuracy of BYTe
by using a significantly improved potential energy surface \citep{jt743},
using empirical energy levels \citep{jt608,jtNH3update}, where available,
to replace computed ones which is of particular importance for high resolution
spectroscopic studies, and improvements in the variational nuclear motion
code TROVE \citep{TROVE,jt626,jt653} made as part of the ExoMol project.
The following sections detail the method used to compute the CoYuTe
line list with particular emphasis on these improvements before discussing
the final line list.

\section{Method}

Construction of a rotation-vibration line list requires three things:
a potential energy surface (PES), dipole moment surface (DMS) and a
nuclear motion program \citep{jt475}. In this work we use the newly
created C2018 PES of \citet{jt743}. Improvements in this potential
were facilited by recent significant progress in assigning ammonia spectra in the near-infrared
\citep{04XuLiYaTr,07LiLeXu.15NH3,08LeLiXu.15NH3,12SuBrHuSc,jt633,jt683} and visible \citep{jt715} regions, as well as the availability of new
hot ammonia spectra \citep{11HaLiBe.NH3,12HaLiBe.NH3,jt616,jt664,jt680}.

As part of this work we constructed a new multireference configuration interaction (MRCI) \citep{88WeKnxx.ai,92KnWexx.ai} DMS using a large, aug-cc-pwCVQZ basis set \cite{89Dunning.ai,92KeDuHa,02PeDuxx.ai}. However,
detailed comparisons with a variety of observations \citep{Coles.thesis} showed that the DMS of \citet{jt466} contracted using the CCSD(T)/aug-cc-pVQZ level of theory used to compute BYTe gave better agreement with observations. The issue here is not so much the level of theory used but the CCSD(T) DMS by \citet{jt466} was generated on a substantially larger grid of points than we could afford for the more computationally-expensive MRCI calculations. Choice of an extended grid is well-known to be crucial in getting a good representation of a DMS \citep{jt511}.

The program TROVE was used to perform the nuclear motion calculations
\citep{TROVE}. TROVE has undergone a number of updates since the construction of BYTe with a particular focus on the production of large line
lists \citep{jt626}. Particularly important is the development of the program GAIN \citep{jt653} which provides a highly
efficient means of computing transition dipoles using GPUs; traditionally this step used to dominate the computer time
requirement but GAIN  speeds the process up by a factor of about one hundred. This meant that for CoYuTe construction and diagonalisation of
the Hamiltonian matrix became the computer-resource limiting step.

Nuclear motion calculations were performed on the Darwin and COSMOS high performance computing (HPC) facilities in Cambridge, UK. At the time
of performing these calculations, each of the computing nodes on the Darwin cluster provided 16 CPUs and a maximum of 64 Gb of RAM, with a wall
clock limit of 36 hours. COSMOS provided 7.3 Gb per CPU and 8 CPUs per node, with a maximum standard job size of 448 Gb and a wall clock
limit of 12 hours. Since multiple nodes can be accessed by a single user at any time, multiple computations could carried out simultaneously.
Our approach to constructing and diagonalising the Hamiltonian matrix for NH$_3$ in TROVE is the same as used by~\citet{jt641} for SO$_3$,
which involves three steps. Firstly the Hamiltonian is calculated and saved to disc. It is then diagonalised separately for each $J$ and
symmetry ($\Gamma_{\rm tot}$) using an MPI-optimized version of the eigensolver PDSYEVD \citep{slug}. Finally, TROVE reads the eigenvectors and eigenvalues and
converts them into a human readable format.

TROVE uses a symmetry-adapted basis set \citep{17YuYaOv.methods}. Construction of the Hamiltonian matrices for each $J$ and $\Gamma_{\rm tot}$ was performed on the COSMOS HPC cluster. In total, for states with
$J=1-43$ and symmetry blocks ($A_2'$, $A_2''$, $E'$, $E''$) this step took 725 hours real-time (11~737 CPU hours), and required a maximum of
223 Gb of RAM for the most expensive calculation, which corresponded to the $E'$ symmetry block of the $J=25$ Hamiltonian. The process was then
moved to Darwin for diagonalisation, which took 272 hours (real-time), and for which the largest matrix to be diagonalised ($J=25$, $E'$ block)
had 246311 rows (see Fig.~\ref{fig:J0converge}) and required the use of 24 parallel nodes.

Evaluation of the line strengths and corresponding Einstein-A coefficients was performed using the GAIN-MPI \citep{jt653} program on the
Wilkes2 GPU cluster at Cambridge. Each GPU node contains 4$\times$Nvidia P100 16GB GPUs. With this program we were able to calculate
approximately 22~000 transitions per second using up to 10 parallel nodes.

\section{Line list construction}

The CoYuTe line list was constructed to cover wavenumbers up to 20~000 \cm, i.e. wavelengths longer than 0.5 $\mu$m for temperatures up to
1500~K. To this end transitions from all states with energies up to 11~000 \cm\ above the ground state were considered which involved
rotational states up to $J=43$; BYTe only includes states with $J \leq 36$. Comparison with the high temperature partition function of
\citet{jt571} suggests that these parameters are more than sufficient to cover temperatures up to 1500~K.

\begin{figure}
1\includegraphics[width=0.45\textwidth]{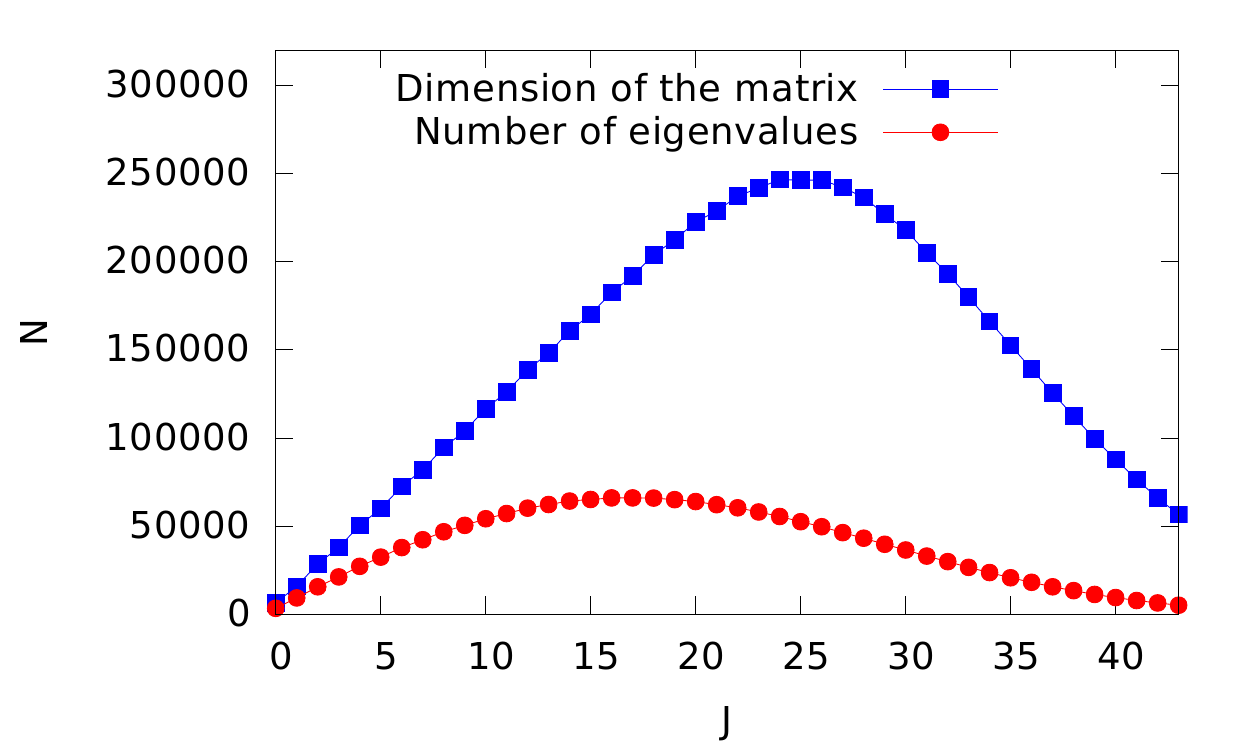}
\includegraphics[width=0.45\textwidth]{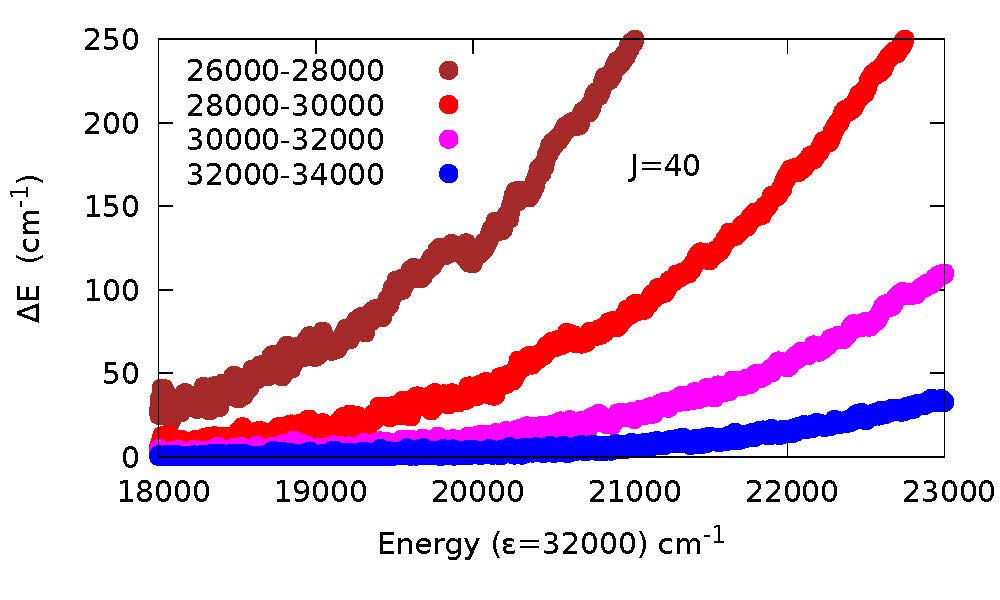}
\caption{Left -- Dimensions of the $E'$-symmetry matrices (squares) and the corresponding number of eigenvalues below 23~000 \cm\ (circles).
Right -- Basis set convergence of~$J=40$~($E'$ symmetry) energies as ($J=0$)-contracted basis set threshold $\epsilon$ is increased from 26~000
to 34~000. The difference $E_{\epsilon=x} - E_{\epsilon=x+2000}$, is displayed for $x=26~000, 28~000, 30~000, 32~000$ vs the energies computed
using $\epsilon=32~000$ \cm.}
\label{fig:J0converge}
\end{figure}

An upper state energy threshold of 23~000 \cm\ was used. This means that a complete representation of the hot spectrum will be obtained for
wavenumbers below 12~000 \cm\ but for wavenumbers above this  value there will be some loss of opacity at higher temperatures. However, as the
C2018 PES used in this work was predominantly tuned to experimental levels up to 7254 \cm\ above the ground state, the precise location of the
higher states included in the calculation is already highly uncertain and further extension to higher energies is hard to justify. The
calculations truncate the $J=0$ contracted basis set at 32~000 \cm, this was necessary to keep within the compute limits available to us and
the consequent reduction in Hamiltonian matrix dimension above $J=25$ is shown in Figure \ref{fig:J0converge}. This truncation means that for high-$J$ states with energies close to 23~000 \cm the CoYuTe energies will not be fully converged. For $J=40$ levels of $E'$ symmetry the resulting error, shown in Fig.~\ref{fig:J0converge}, only becomes significant above approximately 20~000 cm$^{-1}$; for $J=20$ levels the complete energy range is fully converged. It is important to note that this convergence error will affect only a minority of very weak lines that contribute to the continuum at wavelengths approaching, and into, the visible region. We therefore do not expect it to adversely affect the quality of the overall line list.

The number of lines computed is very large so it is desirable to prune the weakest lines. However, experience \citep{jt572} has shown that
including weak lines is important for recovering the correct opacity. To balance these two issues we chose to retain all lines which have an
intensity greater than $1 \times 10^{-36}$ cm$^{-1}$/(molec cm$^{-2}$) at 1500 K. This results in 16.9 billion lines, which is an order of
magnitude more transitions than BYTe, which contains 1.1 billion.

To ensure that the resulting CoYuTe line list provides transition frequencies which are as accurate as possible, we have substituted our
computed energy levels with empirical ones where available. This procedure has been used for other polyatomic ExoMol line lists \citep{jt570,jt734}
and has been shown to give good results \citep{19HuScLe.SO2}. Empirical energy levels were taken from the  MARVEL (measured active
vibration-rotation energy levels) studies of $^{14}$NH$_3$ due to \citet{jt608} and \citet{jtNH3update}. At present 4493 out of 5~095~730 energy levels have been
replaced, and we plan to update this as new experimental data becomes available. Transition wavenumbers between these levels should be highly accurate and, in paricular, suitable for high resolution studies of
(exoplanetary) spectra. We note that at the moment the format used by the ExoMol database \citep{jt548,jt631} does not distinguish between
those transitions which are reproduced with experimental accuracy and those which are the result of theoretical predictions. We are currently
planning an update in the ExoMol database and associated data structures to resolve this problem.

TROVE uses a local mode representation of the vibrational quantum numbers, compared to the more standard normal mode representation. Mapping between these forms is not entirely unambiguous, and particular difficulties arise due to the doubly degenerate modes $\nu_3$ and $\nu_4$. In order to facilitate the assignment of normal mode vibrational quantum labels to our energy levels using purely \textit{ab initio} means, we apply the approach developed by \citet{18ChJeYu} in their treatment of the highly degenerate bending motion of C$_2$H$_2$. Namely, the two 2-dimensional vibrational basis sets associated with $\nu_3$ and $\nu_4$ are transformed into eigenfunctions of the vibrational angular momentum operator squared $\hat{L}^2_z$ using the variational method. Here, $\hat{L}^2_z$ has been used rather than $\hat{L}_z$ as TROVE currently only allows for the evaluation of matrix elements of the quadratic form. The operator $\hat{L}_z^2$ commutes with the (reduced) two-dimensional Hamiltonian operator $\hat{H}^{(2D)}$ for each of the degenerate normal modes $\nu_3$ and $\nu_4$ which are treated as two-dimensional isotropic harmonic oscillators, and so eigenfunctions of $\hat{L}^2_z$ are also eigenfunctions of $\hat{H}^{(2D)}$. The eigenfunctions of $\hat{L}^2_z$ are labelled by their vibrational angular momentum quantum number $L_i=|\ell_i|=\sqrt{\ell_i^2}$ which does not distinguish between positive and negative components of $\ell_i$. For this reason we do not attempt to assign the total vibrational angular momentum $L=|\ell_3+\ell_4|$, and provide only $L_3=|\ell_3|$ and $L_4=|\ell_4|$. The vibrational angular momentum quantum numbers generated from this normal mode representation are mapped onto the local mode representation used by TROVE at the stage of solving the reduced 3D stretching, 2D bending and 1D inversion Schr{\"o}dinger equations (see \citet{17YuYaOv.methods}), and are subsequently propagated through all full-dimensional vibrational and rotational-vibrational calculations.

\section{Results}

The ExoMol database uses a format which separates transitions into a states file (including quantum labels) and a
transitions file \citep{jt548}. Extracts from these two files are given in Tables ~\ref{t:states} and \ref{t:trans}, respectively. These files
themselves can be obtained from\\
\url{ftp://cdsarc.u-strasbg.fr/pub/cats/J/MNRAS/xxx/yy}, or \url{http://cdsarc.u-strasbg.fr/viz-bin/qcat?J/MNRAS//xxx/yy} as well as the ExoMol
website, \url{www.exomol.com}. Updated states files will be made available at \url{www.exomol.com} as and when new empirical energy level data becomes available. In this sense users should consider this version of CoYuTe as living,
whereas the CDS deposits capture the state of database as the point of publication.

For the CoYuTe line list we follow BYTe and give vibrational quantum numbers both in the local mode form
produced by TROVE ($v_1,v_2,v_3,v_4,v_3,v_4$) and the more standard, for ammonia, normal mode form ($n_1,n_2,n_3,n_4,l_3,l_4$). Due to the aforementioned additional step required to map from the local mode to the normal mode representation, the local mode quantum numbers should be regarded as the more
reliable. To distinguish between those energies ($\tilde{E}$) that have been derived from MARVEL and those that have been computed using the C2018 PES we have added an additional column ($E_{CYT}$) to the states file. Where the energy $\tilde{E}$ has been derived from MARVEL, $E_{CYT}$ takes the theoretically determined energy value, otherwise it takes a value of $-1.000000$.

\begin{figure}
\includegraphics[width=0.5\textwidth]{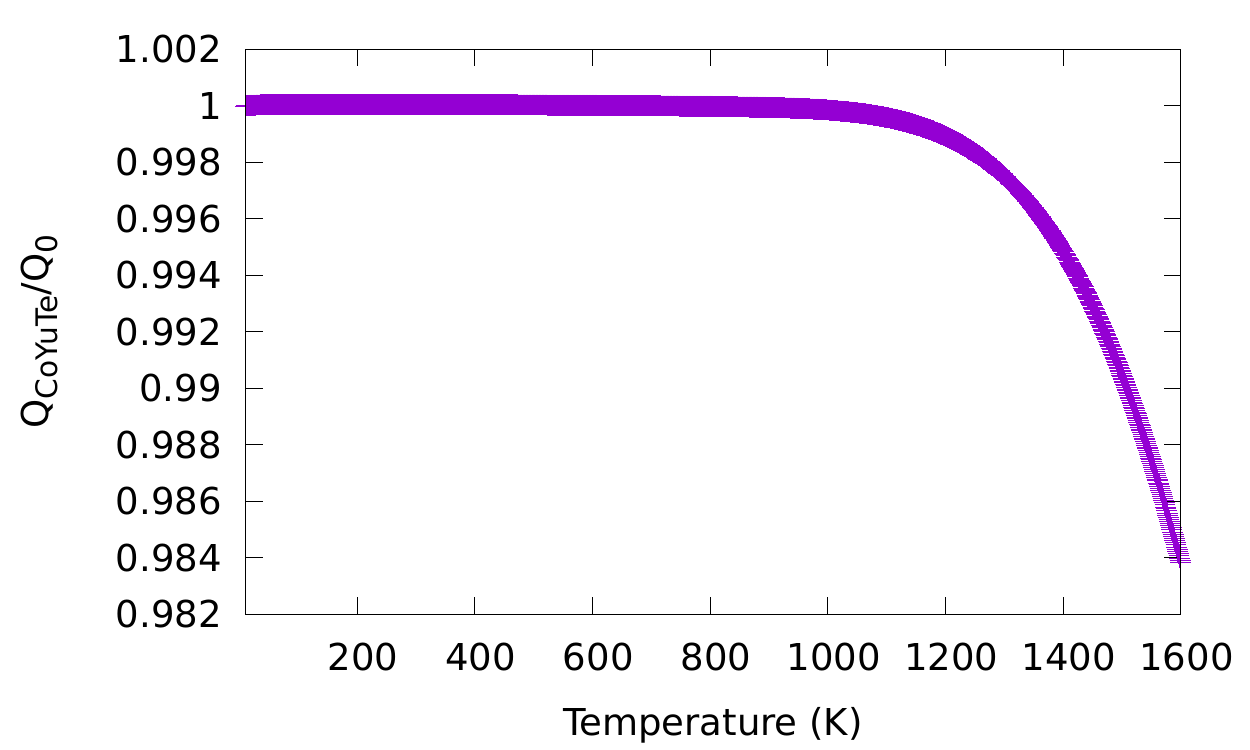}
\caption{Ratio of CoYuTe effective partition function $Q_{\rm CoYuTe}$  to the full partition function of \citet{jt571} ($Q_0$) as a function
of temperature. $Q_{\rm CoYuTe}$ is computed from  the CoYuTe energies with $E_{\rm max} = 11 000$~\cm\ and $J_{\rm max} = 43$.}
\label{fig:pf}
\end{figure}

Partition functions can be used to determine the temperature range over which a line list is complete. \citet{jt181} showed that completeness
as a function of temperature can be quantified by the ratio of the partition sum given by the lower state energy levels to the true partition
sum of the system. \citet{jt571} provided converged partition sums for ammonia which extend to high temperature. Figure~\ref{fig:pf} shows the
ratio of the partition function, $Q(T)$, computed with the CoYuTe energy levels up to the low-energy cut off, to the full partition function of
\citet{jt571}. Below 1200 K this ratio remains close to unity but falls rapidly above this temperature. At 1500 K the ratio is about 0.99 and
the CoYuTe line list can be regarded as effectively complete up to this temperature. It is still possible to use CoYuTe at higher temperatures
but users should be aware that this will increasingly lead to missing opacity.

\begin{figure}
\includegraphics[width=0.8\textwidth]{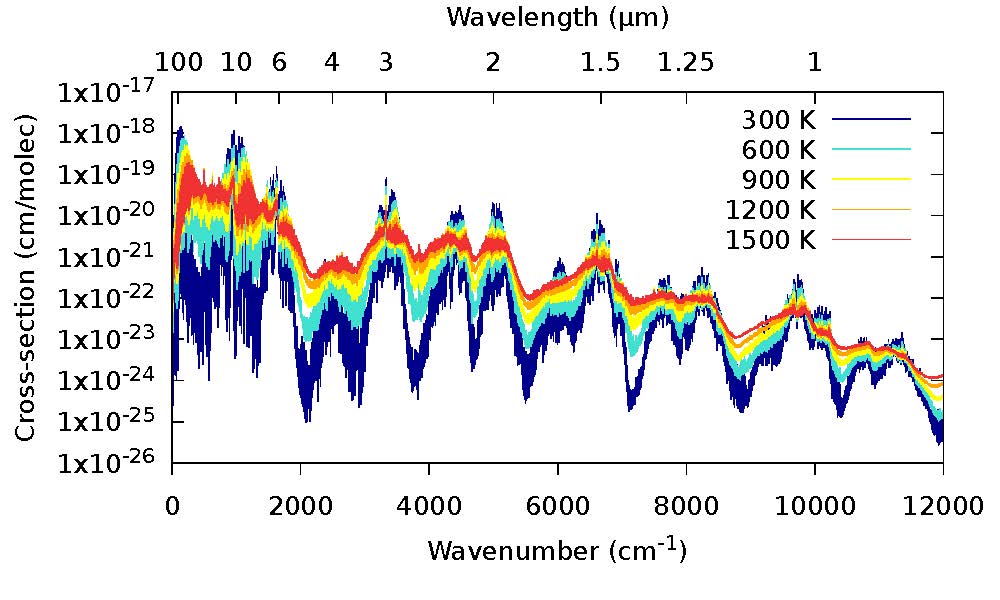}
\caption{Overview of the CoYuTe line list in the 0 -- 12~000 \cm\ region for temperatures up to 1500 K. }
\label{fig:Overview1}
\end{figure}

\begin{figure}
\includegraphics[width=0.8\textwidth]{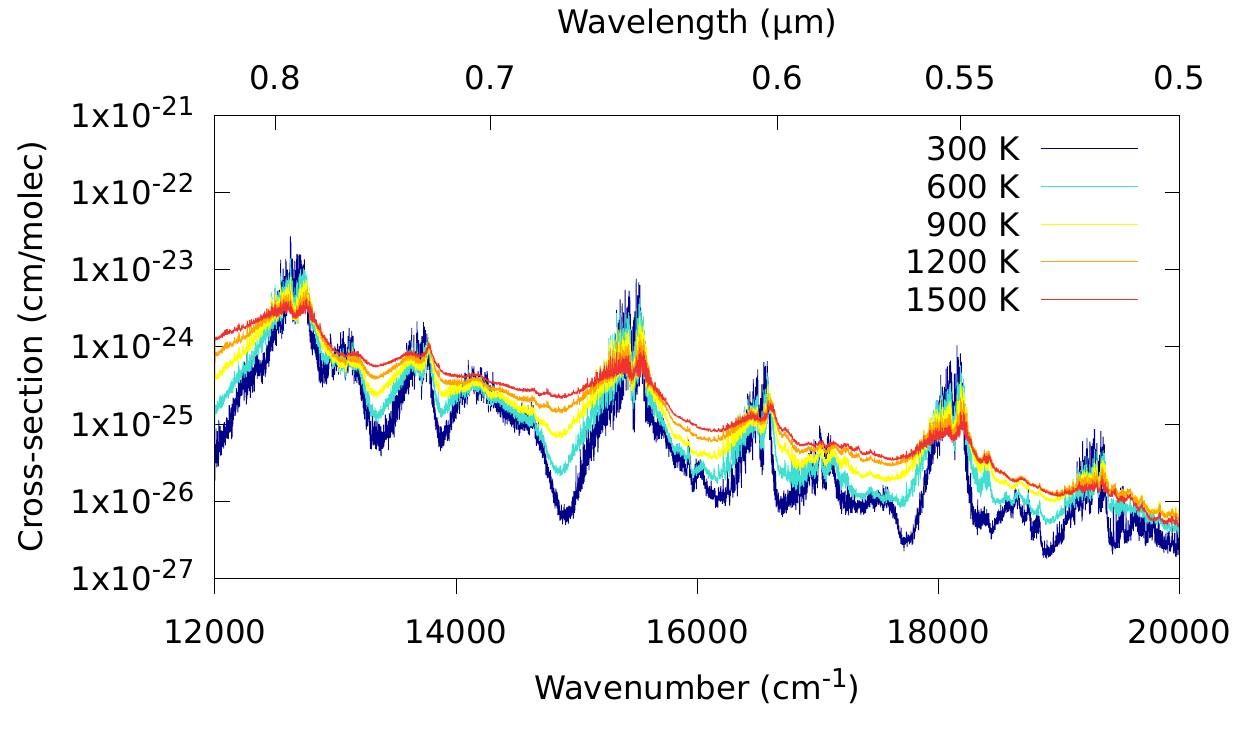}
\caption{Overview of the CoYuTe line list in the 12~000 -- 20~000 \cm\ region for temperatures up to 1500 K.}
\label{fig:Overview2}
\end{figure}

Figures~\ref{fig:Overview1} and ~\ref{fig:Overview2} provide an overview of the CoYuTe line list, which has been split by wavenumber to
reflect that below 12~000 \cm~CoYuTe is expected to be effectively complete, but thereafter will  increasingly suffer from missing opacity
due to the energy thresholds employed. As is usual for such line lists, the total absorption decays approximately exponentially as wavenumber
increases, and the effect of raising the temperature is to reduce the differences between the peaks and troughs in the overall absorption cross
sections.

\begin{landscape}
\begin{table}
\caption{Extracts from the final states file for the CoYuTe line list.}
\begin{tabular}{rrccccccccccccccccccccrcr}
\hline\hline
$i$ &  $\tilde{E}$  & \gtot  & $J$ & $p$ & $\Gamma$& $N_b$& $n_1$ & $n_2$ &$n_3$& $n_4$ &$l_3$& $l_4$& $\tau_i$& $J$& $K$ & $\tau_r$
& $v_1$ &$v_2$ &$v_3$& $v_4$& $v_5$ &$v_6$& $\Gamma_{\rm vib}$& $E_{CYT}$\\
\hline
7779	&	0.793374	&	12	&	0	&	-	&	5	&	1	&	0	&	0	&	0	&	0	&	0	&	0	&	1	&	0	&	0	&	0	&	0	&	0	&	0	&	0	&	0	&	1	&	5	&	0.793374	\\			
7780	&	968.121906	&	12	&	0	&	-	&	5	&	2	&	0	&	1	&	0	&	0	&	0	&	0	&	1	&	0	&	0	&	0	&	0	&	0	&	0	&	0	&	0	&	3	&	5	&	968.121906	\\			
7781	&	1882.177407	&	12	&	0	&	-	&	5	&	3	&	0	&	2	&	0	&	0	&	0	&	0	&	1	&	0	&	0	&	0	&	0	&	0	&	0	&	0	&	0	&	5	&	5	&	1882.177426	\\			
7782	&	2895.521390	&	12	&	0	&	-	&	5	&	4	&	0	&	3	&	0	&	0	&	0	&	0	&	1	&	0	&	0	&	0	&	0	&	0	&	0	&	0	&	0	&	7	&	5	&	2895.521820	\\			
7783	&	3217.582251	&	12	&	0	&	-	&	5	&	5	&	0	&	0	&	0	&	2	&	0	&	0	&	1	&	0	&	0	&	0	&	0	&	0	&	0	&	0	&	2	&	1	&	5	&	3217.579464	\\			
7784	&	3337.097095	&	12	&	0	&	-	&	5	&	6	&	1	&	0	&	0	&	0	&	0	&	0	&	1	&	0	&	0	&	0	&	0	&	1	&	0	&	0	&	0	&	1	&	5	&	3337.097117	\\			
7785	&	4061.640928	&	12	&	0	&	-	&	5	&	7	&	0	&	4	&	0	&	0	&	0	&	0	&	1	&	0	&	0	&	0	&	0	&	0	&	0	&	0	&	0	&	9	&	5	&	-1.000000	\\			
7786	&	4173.090304	&	12	&	0	&	-	&	5	&	8	&	0	&	1	&	0	&	2	&	0	&	0	&	1	&	0	&	0	&	0	&	0	&	0	&	0	&	0	&	2	&	3	&	5	&	-1.000000	\\			
7787	&	4320.030627	&	12	&	0	&	-	&	5	&	9	&	1	&	1	&	0	&	0	&	0	&	0	&	1	&	0	&	0	&	0	&	0	&	1	&	0	&	0	&	0	&	3	&	5	&	4320.030627	\\			
7788	&	4843.355303	&	12	&	0	&	-	&	5	&	10	&	0	&	0	&	0	&	3	&	0	&	3	&	1	&	0	&	0	&	0	&	0	&	0	&	0	&	0	&	3	&	1	&	5	&	-1.000000	\\			

\hline\hline
\end{tabular}
\label{t:states}
\mbox{}\\

{\flushleft
$i$:   State counting number.     \\
$\tilde{E}$: State energy in \cm. \\
\gtot: Total state degeneracy.\\
$J$: Total angular momentum.            \\
$p$:     Total parity of the state.\\
$\Gamma$:   Total symmetry index in D$_{\rm 3h}$(M)\\
$N_b$:   Counting number in the $(J,p,\Gamma)$ block.\\
$n_1$:      Symmetric stretch quantum number (normal mode).\\
$n_2$:     Symmetric bend quantum number (normal mode).\\
$n_3$:      Asymmetric stretch quantum number (normal mode).\\
$n_4$:       Asymmetric bend quantum number (normal mode).\\
$l_3$:    Asymmetric stretch vibrational angular momentum quantum number (normal mode).\\
$l_4$:     Asymmetric bend  vibrational angular momentum quantum number (normal mode).\\
$\tau_i$:   Inversion parity (0 or 1).\\
$J$: Total angular momentum.            \\
$K$:       Projection of $J$ on molecular symmetry axis.\\
$\tau_r$:  Rotational parity (0 or 1).\\
$v_1$:   Local mode vibrational quantum number.\\
$v_2$:   Local mode vibrational quantum number.\\
$v_3$:   Local mode vibrational quantum number.\\
$v_4$:   Local mode vibrational quantum number.\\
$v_5$:   Local mode vibrational quantum number.\\
$v_6$:   Local mode vibrational quantum number.\\
$\Gamma_v$: Vibrational symmetry (local mode).\\
$E_{CYT}$: Theoretical state energy in \cm. \\
}
\end{table}
\end{landscape}

\begin{table}
\caption{Extract from the transitions file for the CoYuTe line list. }
\begin{tabular}{rrrr}
\hline\hline
$f$ & $i$ & $A_{fi}$ & $\nu_{if}$\\
\hline
2864780	&	2768903	&	2.69E-30	&	0.000001	\\
4664622	&	4624800	&	2.06E-25	&	0.000001	\\
4664622	&	4624800	&	2.06E-25	&	0.000001	\\
1785731	&	1883312	&	4.64E-25	&	0.000009	\\
1225073	&	1315315	&	1.86E-22	&	0.000014	\\
4595123	&	4550448	&	2.97E-21	&	0.000034	\\
1883390	&	1981447	&	2.24E-26	&	0.000036	\\
4751546	&	4716239	&	1.24E-20	&	0.000046	\\
4400507	&	4345966	&	1.33E-21	&	0.000047	\\
3866874	&	3635182	&	1.20E-23	&	0.000060	\\

\hline\hline
\end{tabular}
\label{t:trans}
\mbox{}\\
{$f$}: Upper state counting number.  \\
{$i$}: Lower state counting number. \\
$A_{fi}$: Einstein-A coefficient in s$^{-1}$.\\
\end{table}

%


%

\section{Validation}

\begin{figure}
\includegraphics[width=0.45\textwidth]{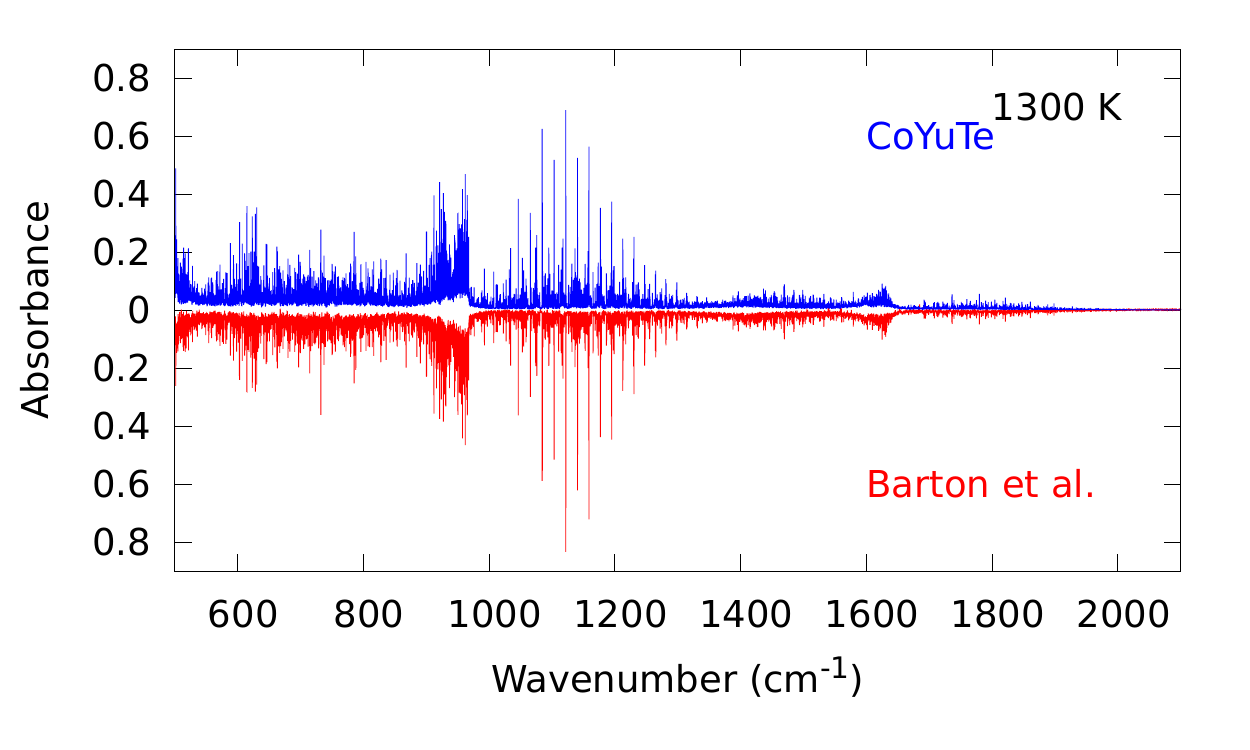}
\includegraphics[width=0.45\textwidth]{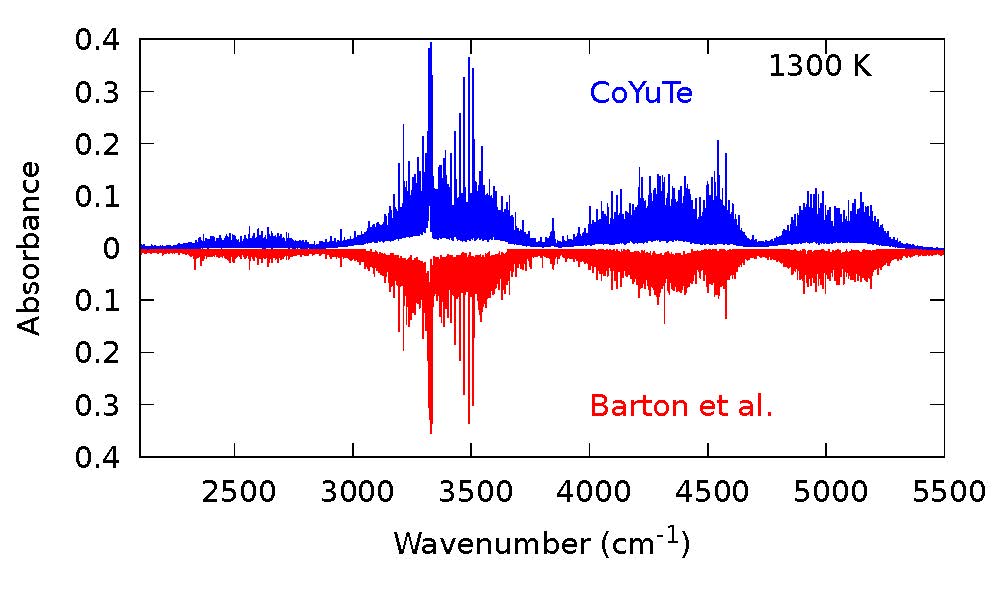}
\caption{Synthetic absorbance spectra computed using CoYuTe compared to the measurements by \citet{jt616,jt664} at a temperature of 1300 K.}
\label{fig:Bart}
\end{figure}

\begin{figure}
\includegraphics[width=0.33\textwidth]{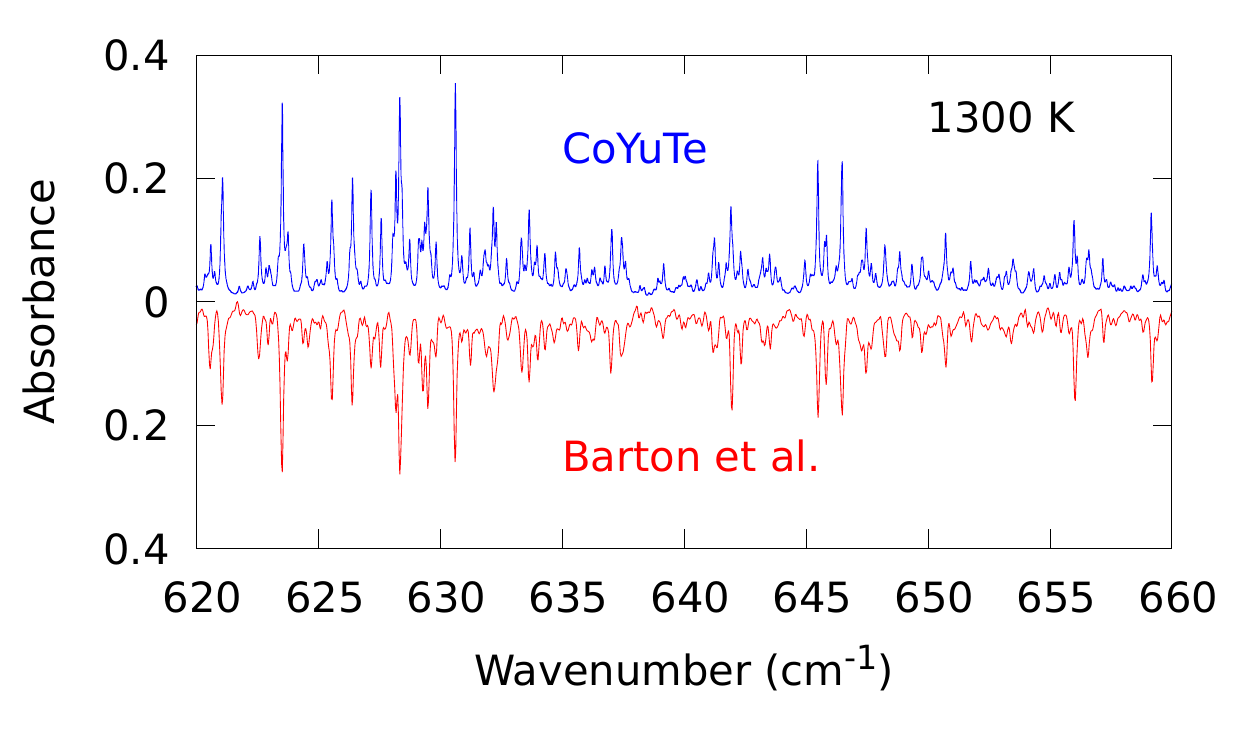}
\includegraphics[width=0.33\textwidth]{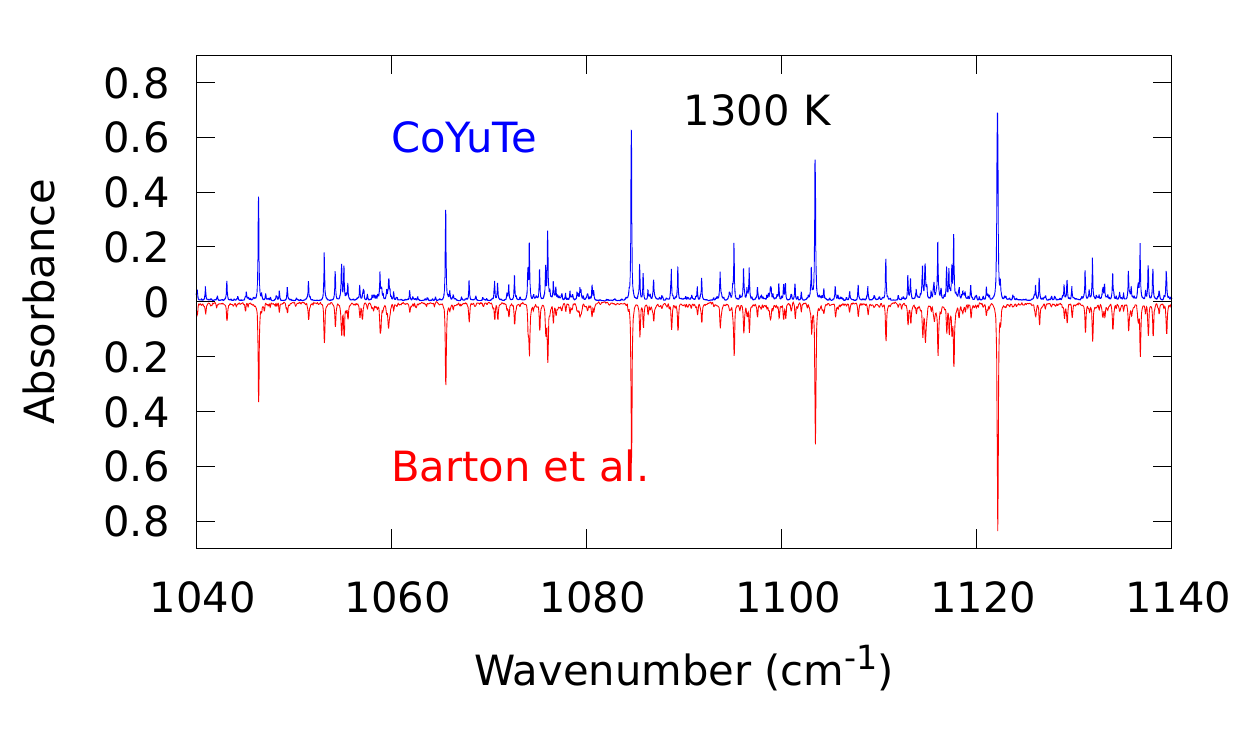}
\includegraphics[width=0.33\textwidth]{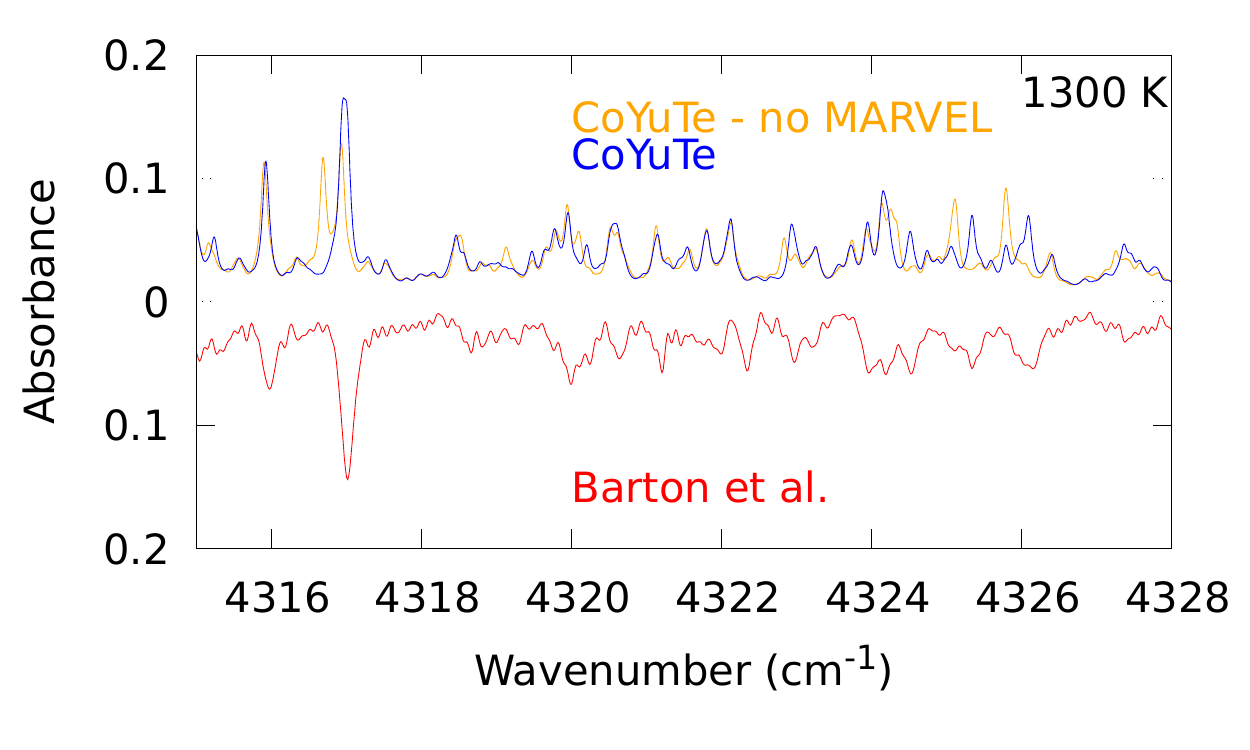}
\caption{Close-up comparison of the synthetic absorbance spectra computed using CoYuTe, and the measurements by \citet{jt616,jt664} at a temperature of 1300 K.}
\label{fig:Bartclose}
\end{figure}

Since the production of BYTe, there have been several experimental studies focusing on high temperature ammonia spectra in the near infrared.
To validate the application of CoYuTe to high temperature ammonia studies we performed a systematic comparison of the CoYuTe predictions with
laboratory measurements from several of these sources.

\begin{figure}
\includegraphics[width=0.45\textwidth]{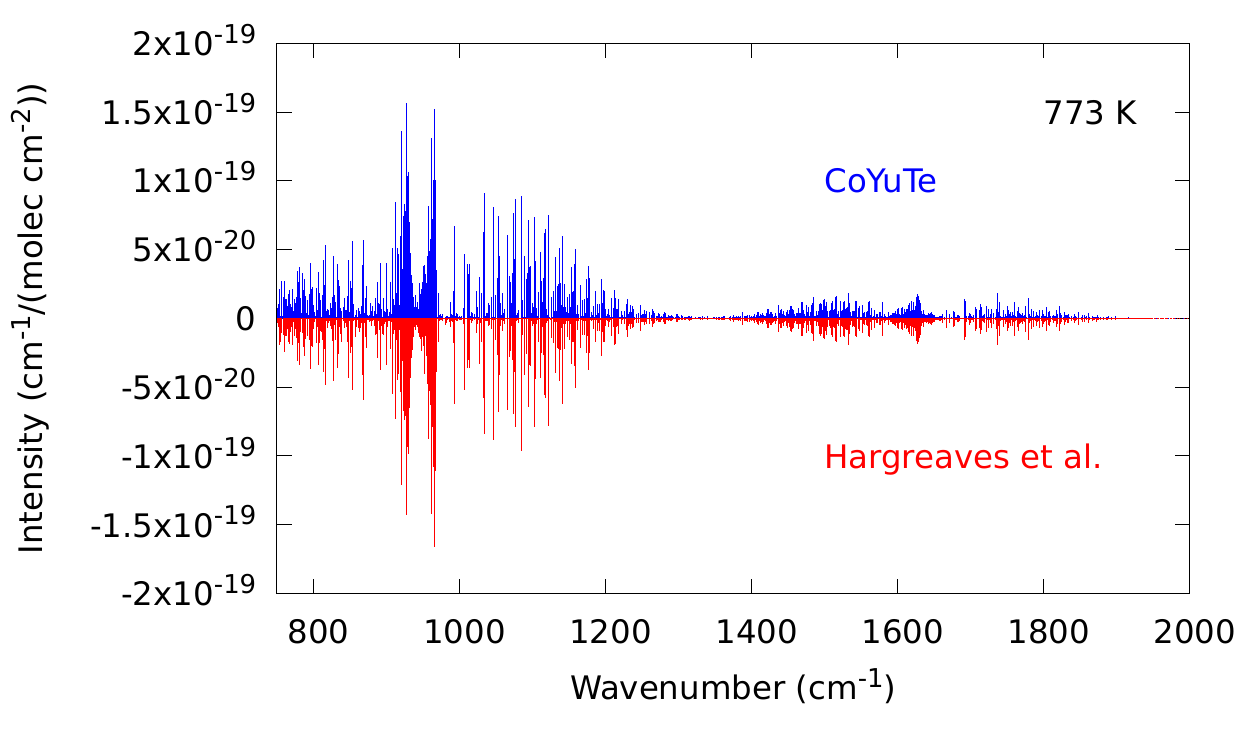}
\includegraphics[width=0.45\textwidth]{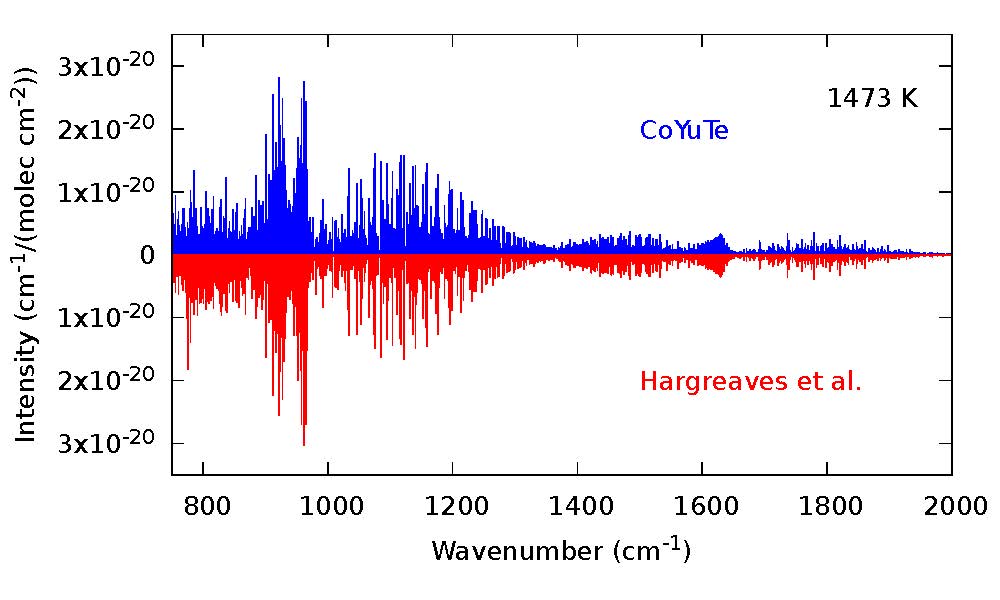}
\caption{Comparison of the ammonia line lists measured by  \citet{11HaLiBe.NH3} with the CoYuTe predictions for temperatures of
773~K and 1473~K.}
\label{fig:Harg}
\end{figure}

\begin{figure}
\includegraphics[width=0.45\textwidth]{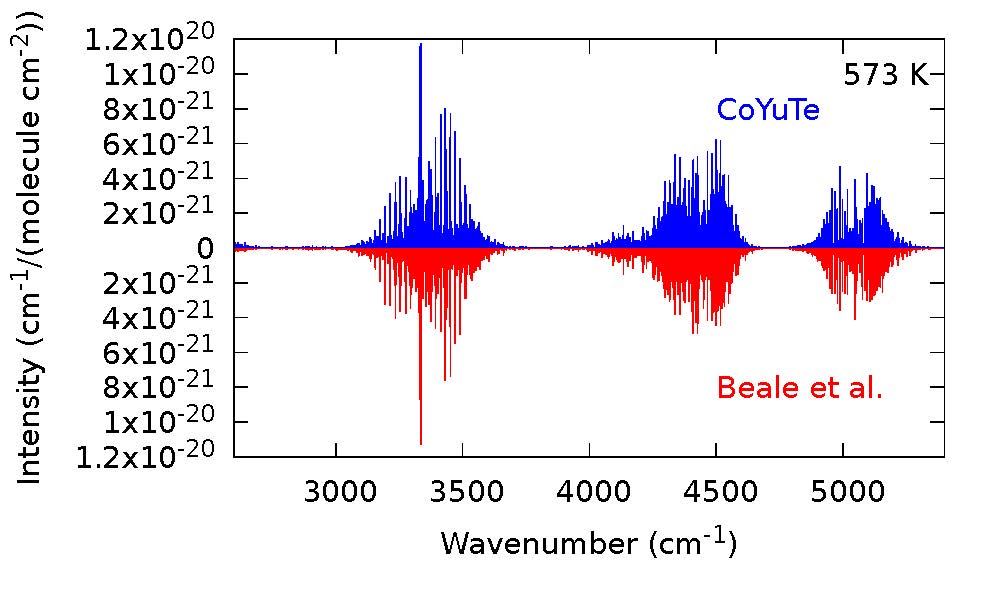}
\includegraphics[width=0.45\textwidth]{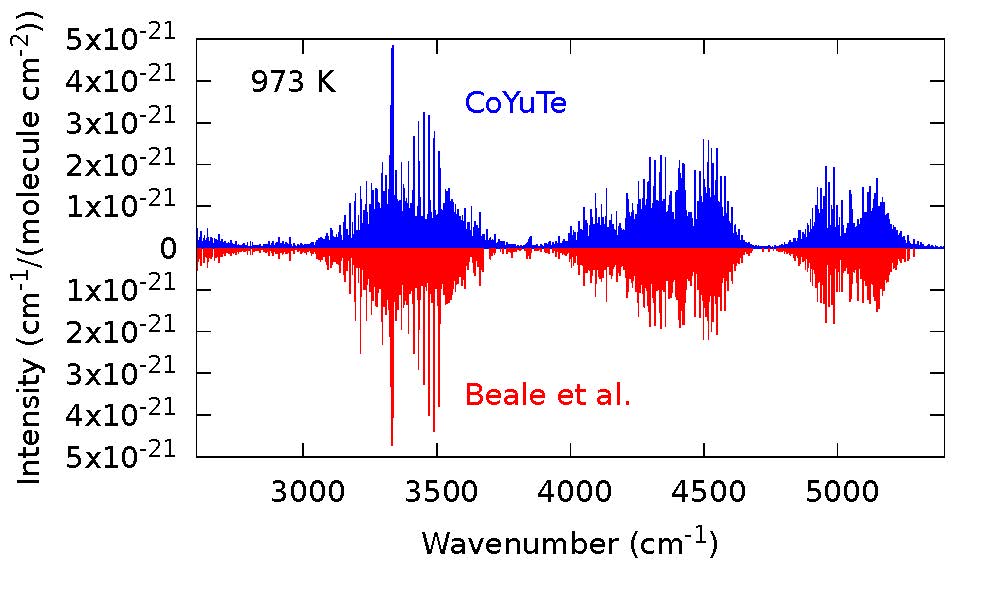}
\caption{Comparison of the ammonia line lists measured by \citet{jt680} with the CoYuTe predictions for temperatures of
573~K and 973~K.}
\label{fig:Beale}
\end{figure}

\begin{figure}
\includegraphics[width=0.33\textwidth]{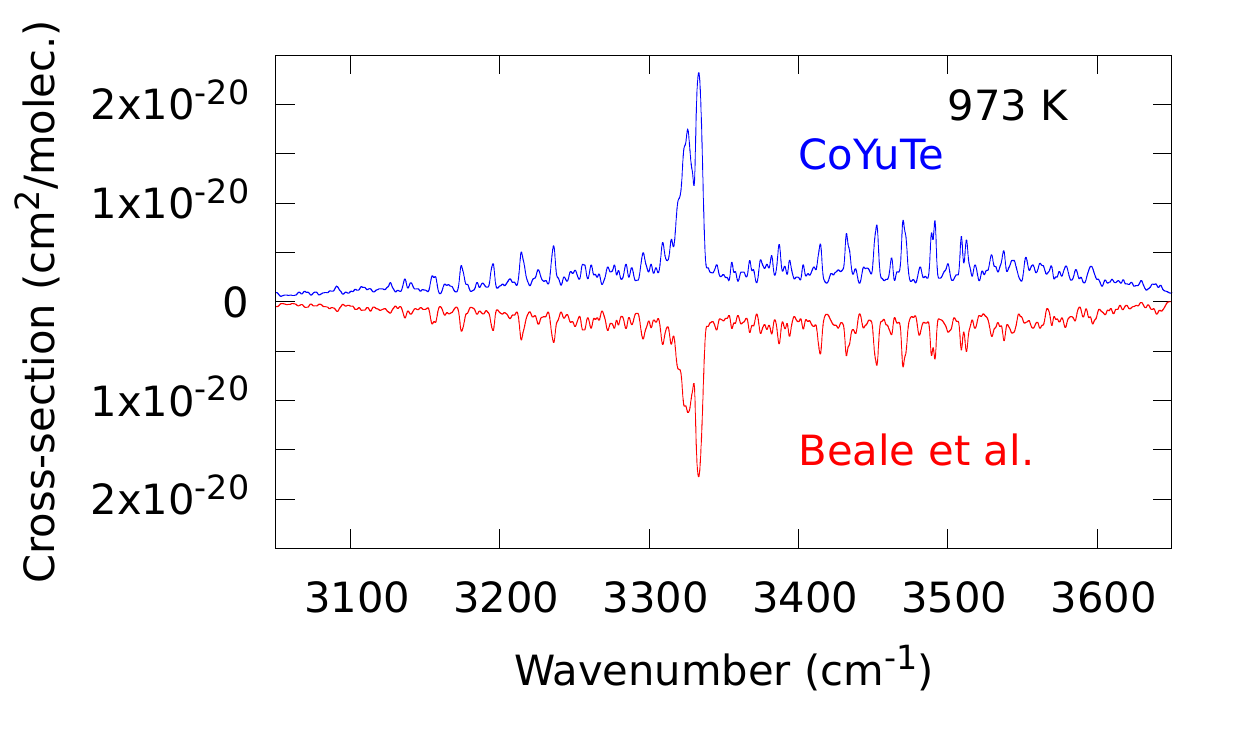}
\includegraphics[width=0.33\textwidth]{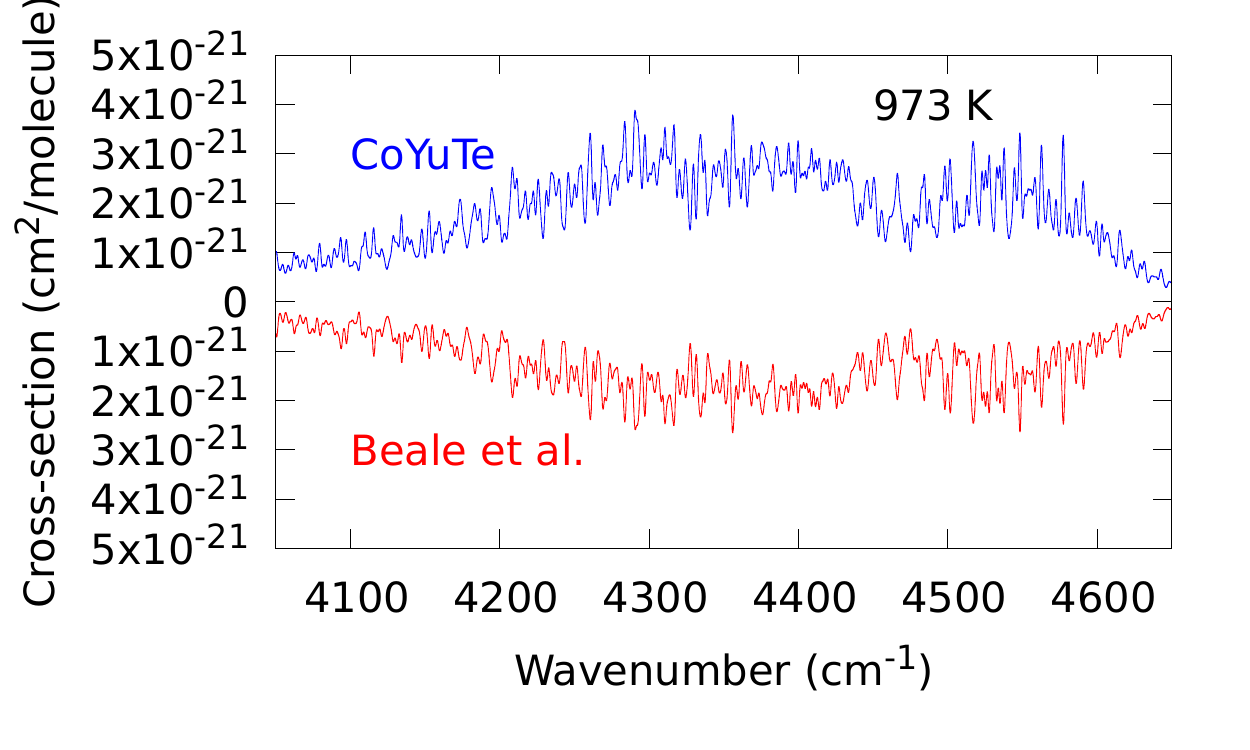}
\includegraphics[width=0.33\textwidth]{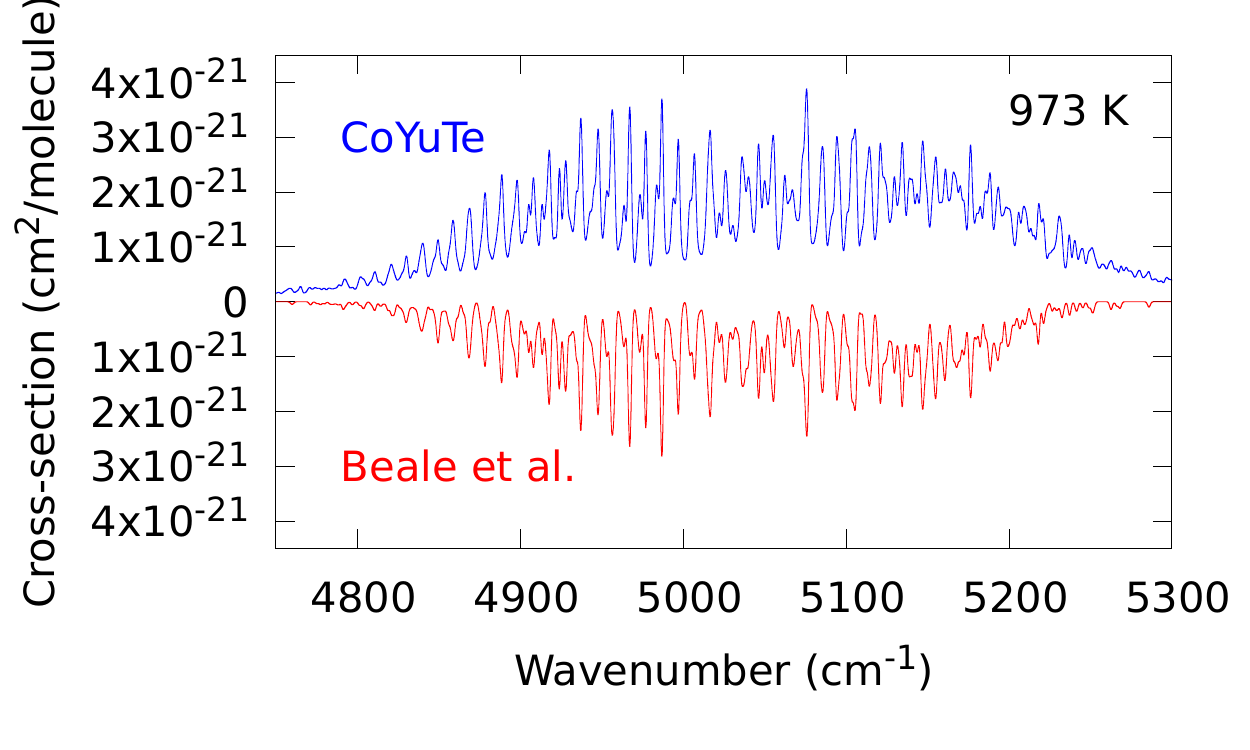}
\caption{Close-up comparison of synthetic spectra generated using the experimental line list by \citet{jt680} and the CoYuTe line list for a temperature of 973 K. Both line lists have been convoluted with Gaussian profiles with HWHM $=1.0$ cm$^{-1}$.}
\label{fig:Bealeclose}
\end{figure}


\begin{figure}
\includegraphics[width=0.5\textwidth]{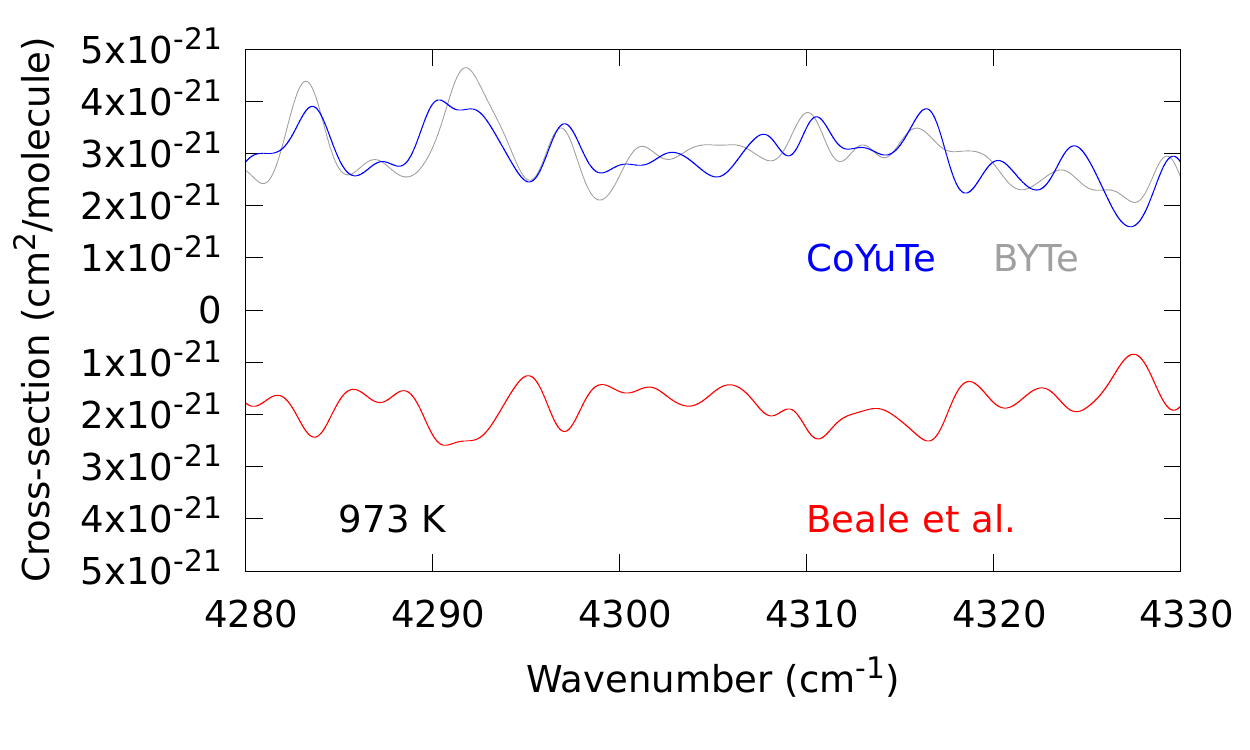}
\includegraphics[width=0.5\textwidth]{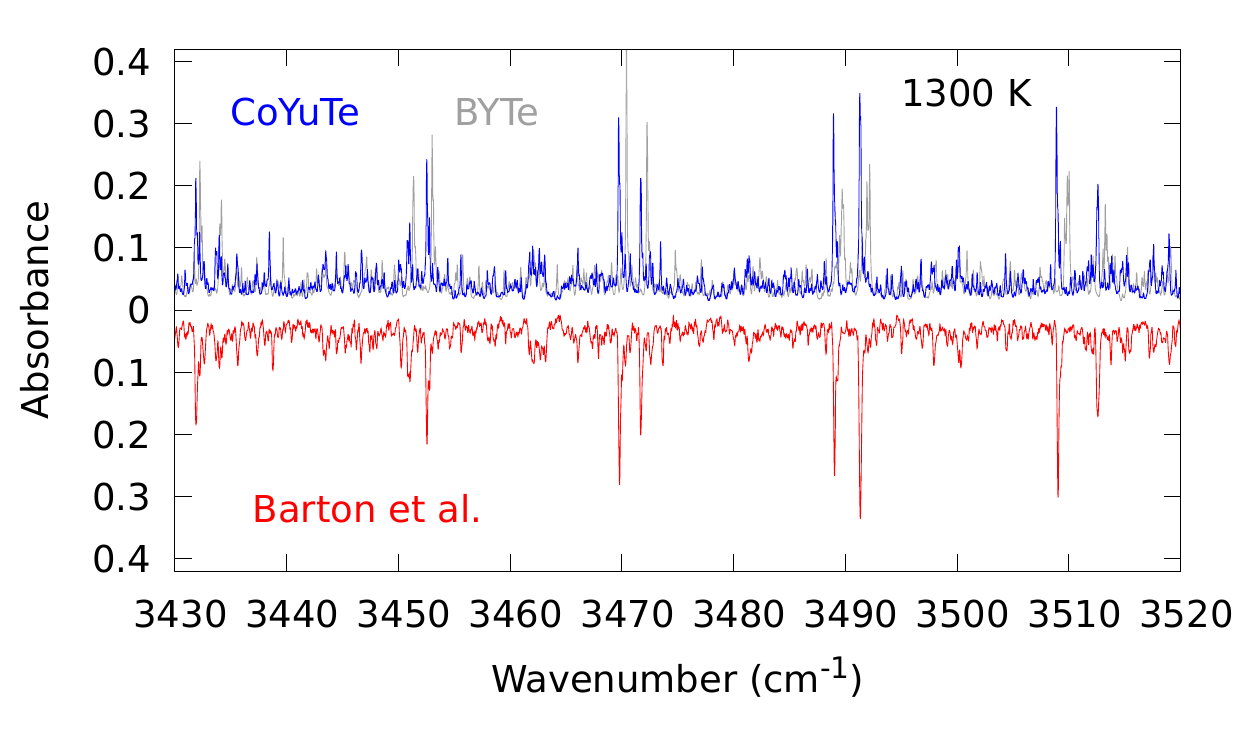}
\caption{Synthetic absorption spectra generated using BYTe and CoYuTe compared to the measurements by \citet{jt664} and the cross-sections (Gaussian profile, HWHM$=1.0$ cm$^{-1}$) calculated using the experimental line lists by \citet{jt680}.}
\label{fig:CoYuTe-BYTe}
\end{figure}

\citet{jt616,jt664} recorded moderate resolution (0.09 \cm) ammonia spectra in the 500--2100 \cm~ and 2100--5500 \cm~ regions for temperatures up
to 1300 K and atmospheric pressure using Fourier transform infrared (FTIR) spectroscopy. They provide their measured absorbance spectra,
partially assigned peak lists measured at 1300 K, and accompanying experimentally derived upper state energies. Figures \ref{fig:Bart} and
\ref{fig:Bartclose} compare synthetic absorbance spectra calculated using the CoYuTe line list to their measured spectra. For these
comparisons, CoYuTe cross sections were computed using the ExoCross program \citep{jt708} and Voigt profiles with halfwidth half maximum (HWHM)
of 0.09 \cm. These were then converted into spectral absorbance using the procedure outlined by \citet{jt616}. Overall agreement is very good,
there are some discrepancies in magnitude for a number of strong absorbance peaks, which are predominantly due to small inaccuracies in CoYuTe
transition frequencies which result in incorrect blending or separation of lines. Replacement of the CoYuTe energies computed from first
principles with the empirically derived energies from MARVEL resolve a number of these cases which would otherwise have been present in the
line list, as is illustrated in the rightmost panel of Figure \ref{fig:Bartclose}.

\citet{11HaLiBe.NH3,12HaLiBe.NH3} and \citet{jt680} used high resolution (0.01 \cm) Fourier transform infrared emission spectra to
produce line lists for hot ammonia in the 740--2100, 1650--4000 and  2400--5500 \cm\ regions respectively. Figure~\ref{fig:Harg} makes a
comparison between \citet{11HaLiBe.NH3} and CoYuTe for temperatures of 773 K and 1473 K. Figure \ref{fig:Beale} shows a similar comparison for
the work by \citet{jt680}. There are a number of instances at 973~K where CoYuTe lines seemingly underestimate line intensities compared to those of
Beale et al. This is because, despite the high resolution nature of their spectra, a number of blended lines have gone unresolved in
the experimental spectrum, resulting in two transitions in close proximity being perceived as one doubly strong transition.
 In this respect it is important to note the excellent agreement between absorption cross-sections (Gaussian profile, HWHM 1.0 cm$^{-1}$) calculated using the
CoYuTe line list and that by \citet{jt680}, which is shown in Figure \ref{fig:Bealeclose}. Referring back to the absorbance spectra measured by \citet{jt664}, shown in Figs. \ref{fig:Bart} and \ref{fig:Bartclose}, it is clear that \citet{jt680} are missing significant opacity.

CoYuTe is the successor to BYTe, and aside from containing more opacity and extended coverage, the line positions and line intensities are significantly more accurate. Detailed analysis of the energy level predictions and room temperature spectra computed using the C2018 PES compared to that of BYTe has already been performed by \cite{jt743}, with a further update by \cite{Coles.thesis}. In light of the astronomical applications intended for CoYuTe it is important to illustrate these improvements at high temperatures as well. Figure \ref{fig:CoYuTe-BYTe} shows synthetic hot spectra generated using BYTe and CoYuTe compared to the absorbance spectra measured by \citet{jt664} and cross-sections (Gaussian profile, HWHM 1 cm$^{-1}$) calculated using the experimental line lists by \citet{jt680}. Clearly both experimental line positions and line intensities are substantially better represented by CoYuTe than they are BYTe. It is important to note that both BYTe and CoYuTe utilise the same DMS, and so the improvement in many line intensities is solely due to the improved PES which is inexorably linked to the linestrength through the wavefunctions.

\begin{figure}
\includegraphics[width=0.5\textwidth]{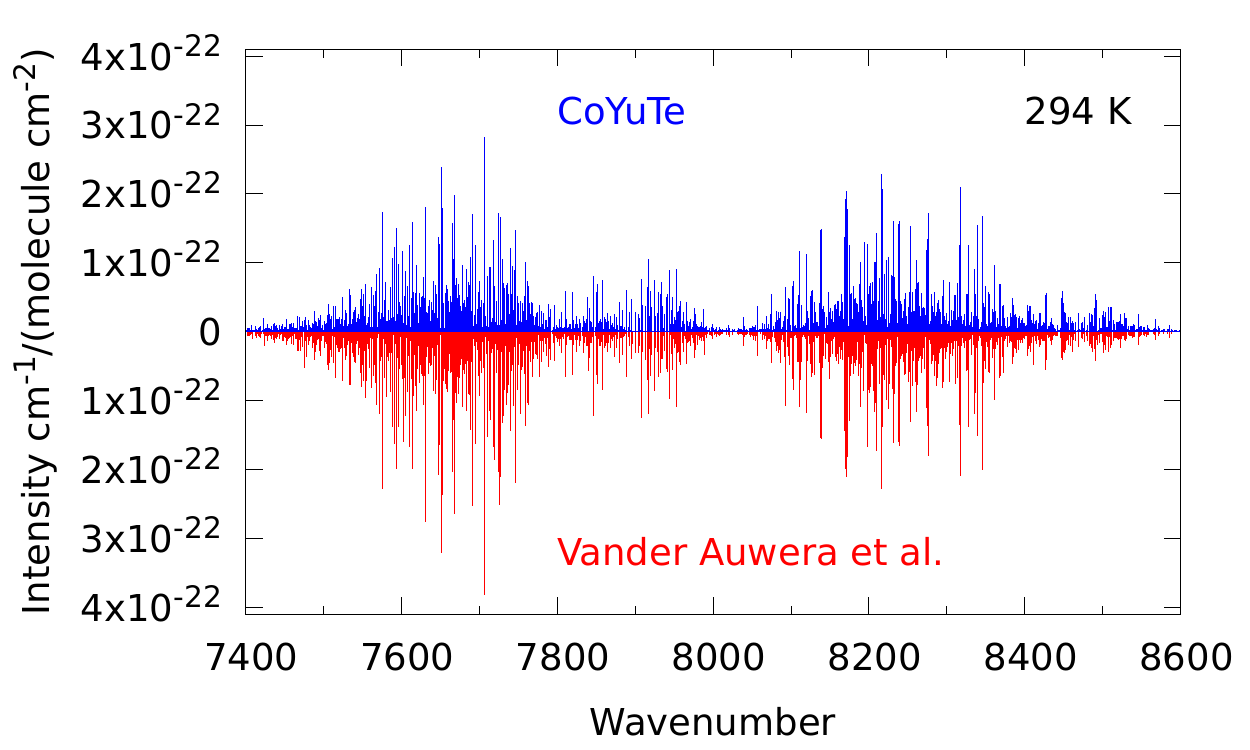}
\includegraphics[width=0.5\textwidth]{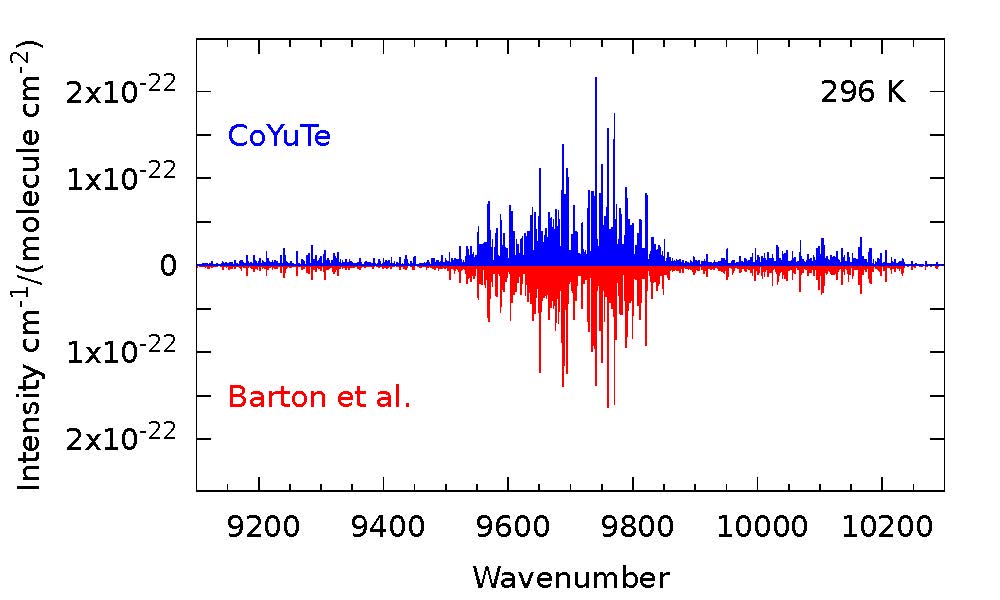}
\caption{Comparison of the simulated CoYuTe and observed spectra of NH$_3$ at T = 294 K for the 7400--8600 region \citep{18VaVaxx.NH3}, and at T = 296 K for 9100--10300 \cm region \citep{jt683}}
\label{fig:VaVa-Bart2}
\end{figure}


Above 5300 \cm, to our knowledge, no laboratory measurements of hot ammonia exist in the literature. Therefore our comparisons with laboratory data at higher wavenumbers are restricted to room temperature only. Recently \citet{jt633,jt683} produced partially assigned line lists in the 7400--8600 \cm and 9000--10400 regions through analysis of FTIR spectra recorded in 1980 at the Kitt Peak National Observatory. \citet{18VaVaxx.NH3} later re-analysed the same  spectra between 7400 and 8600 \cm~using an improved procedure, re-measured and re-analysed the region using an improved experimental setup, and produced new line lists from their analyses. Figure \ref{fig:VaVa-Bart2} (left panel) compares simulated stick spectra calculated using the CoYuTe line list to the work by \citet{18VaVaxx.NH3}. Overall absorption features are represented well, however, there is a tendency to underestimate some bands by as much as 30\%. This underestimation is a feature of our chosen DMS, and has been discussed previously by \citet{jt743}. Figure \ref{fig:VaVa-Bart2} (right panel) presents a similar comparison for the work by \citet{jt683}. Once again the overall band structure is represented well, and the reader is directed to \citet{jt743} for a more detailed analysis.

\begin{figure}
\includegraphics[width=0.5\textwidth]{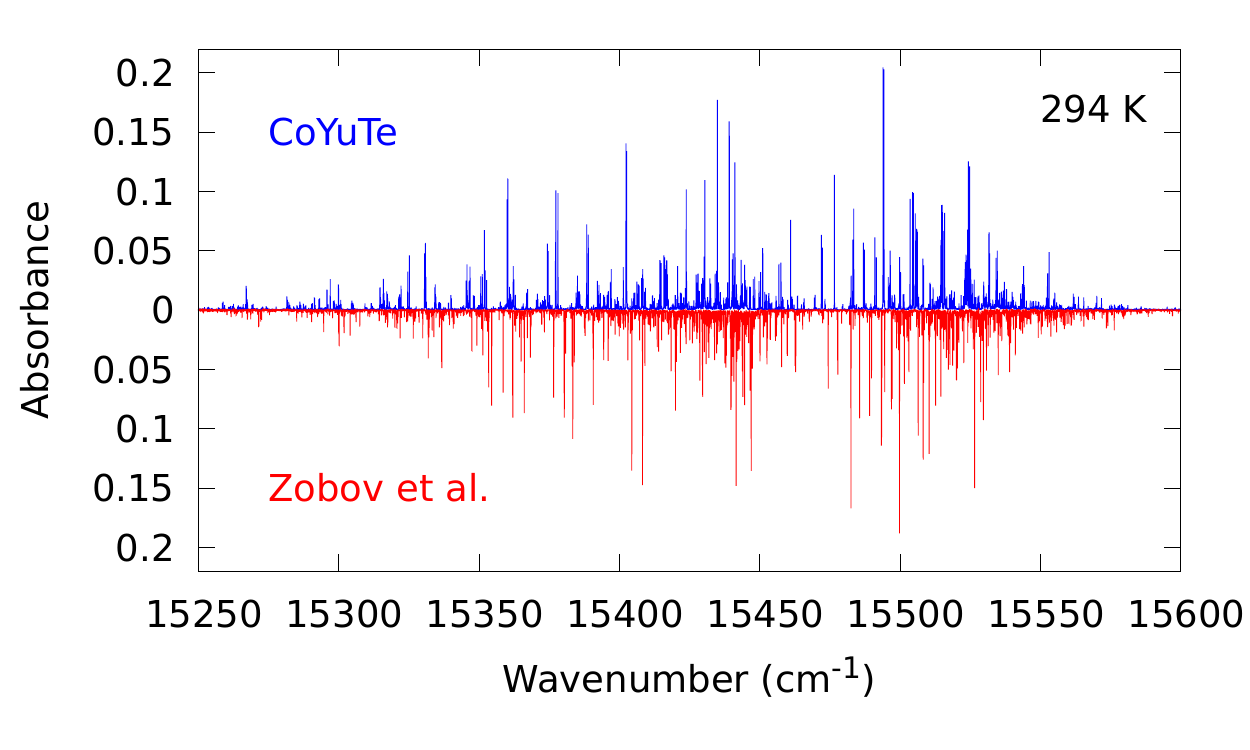}
\includegraphics[width=0.5\textwidth]{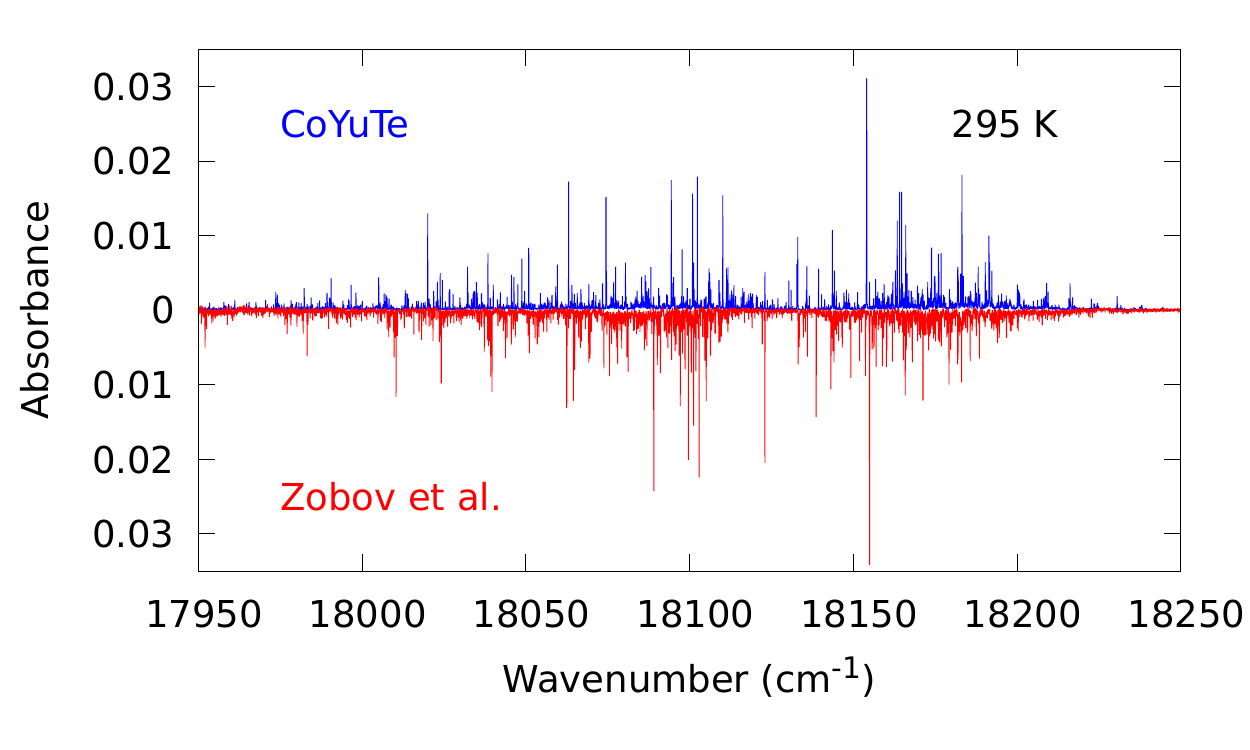}
\caption{Comparison of the simulated CoYuTe and observed \citep{jt715}
spectra of NH$_3$ at T = 294/295 K for the 15250--15600 and 17950--18250 \cm\ regions}
\label{fig:Zobov}
\end{figure}

Several studies have also focussed on the measurement of ammonia spectra at visible wavelengths. The most recent and comprehensive of these is the analysis by \citet{jt715} of a high resolution Kitt Peak spectrum recorded in 1980. Their measured room temperature absorbance spectra is compared with synthetic spectra generated using CoYuTe at red (15~500 \cm) and green (18~000 \cm) visible frequencies in Figure \ref{fig:Zobov}. For these comparisons, CoYuTe cross sections were computed using the ExoCross program \citep{jt708} and Voigt profiles with HWHM
of 0.01 \cm. No information regarding the ammonia concentration in the Kitt Peak sample gas could be found, so we used concentration of 0.7\% (red spectrum) and 0.9\% (green spectrum) for our absorbance calculations, as these approximately matched the peak heights of our calculated spectra to those of \citet{jt715}. In this regard the comparisons presented in Figure \ref{fig:Zobov} should not be taken as evidence of the accuracy of our absolute line intensities at visible wavelengths. However, some degree of reliability has already been confirmed in the work by \citet{jt745}. Regarding relative intensities the overall band profile of the red spectrum (Figure \ref{fig:Zobov}, left) shows reasonable agreement between the calculated and measured spectra, although in most cases it is not possible to match individual lines by eye. In this region \citet{jt715} noted discrepancies between the Kitt Peak measurements and the preliminary version of CoYuTe used in their analysis of up to 6 \cm\ for $J=1-7$ lines, and we expect a similar level of error here.
The green spectrum (Figure \ref{fig:Zobov}, right) displays substantially worse agreement in overall structure, which is unsurprising seeing as only the 6$\nu_{\rm NH}$ stretching band centres were included in the refinement of our potential \citep{jt743}. Inclusion of these band centres in the refinement procedure acts to offset the convergence error due to our vibrational stretching basis, but does not account for inaccuracies associated with rotational excitations within these bands. For this reason it is not recommended to use CoYuTe for high resolution studies at short wavelengths in the visible region. Certainly additional assignments of the 5$\nu_{\rm NH}$ and 6$\nu_{\rm NH}$ stretching overtones would help rectify these errors in the next generation of theoretical line lists.

\begin{figure}
\centering
\includegraphics[width=0.5\textwidth]{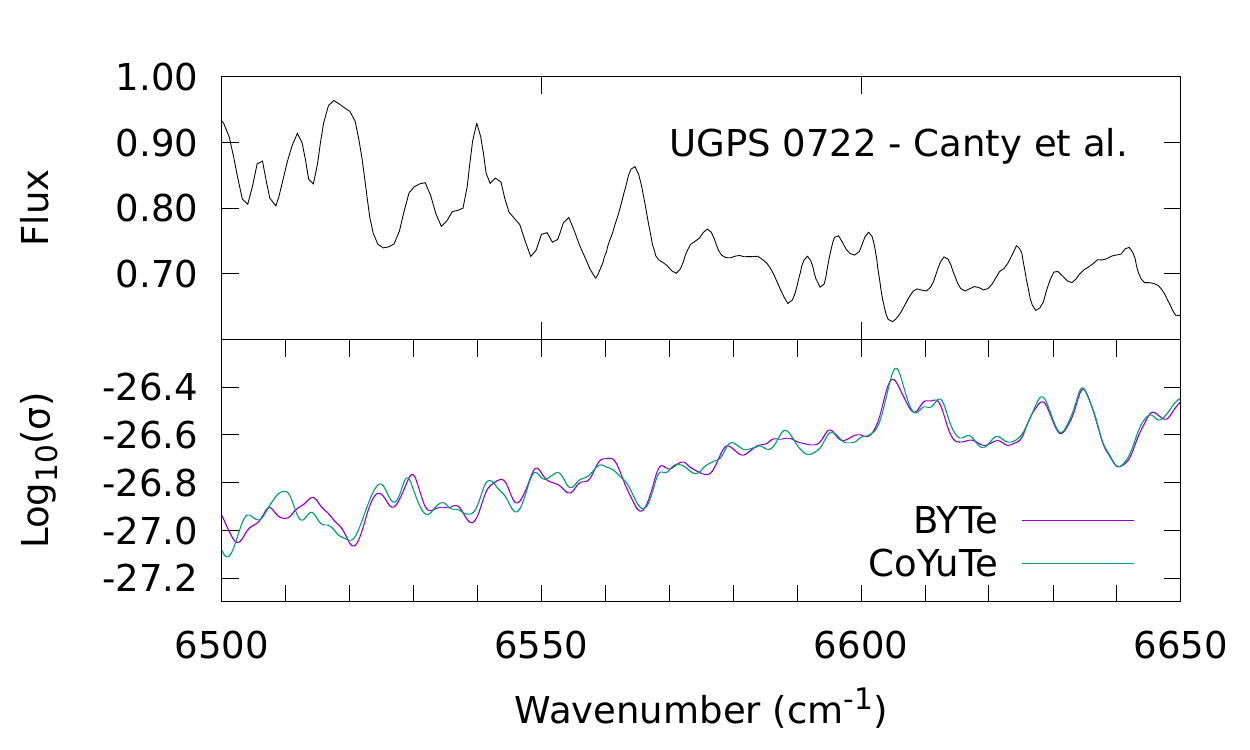}
\caption{Comparison of the the spectrum of late T dwarf  UGPS 0722, which
has an effective temperature of 500 K, due to \citet{jt596}. Our models
assume water, methane, ammonia atmosphere with the only difference being
the line list used to represent ammonia. Results
are presented as
$\log10$ of the cross-sections, $\sigma$ given in
in cm$^2$/molecule; see text for details.}
\label{fig:Canty}
\end{figure}

Finally there are a few high resolution astronomical spectra of hot objects containing
ammonia. One example are spectra of brown dwarf stars. \citet{jt596}
analysed the spectra of late T dwarfs assigning a variety of features to
ammonia and methane using the BYTe and 10to10 \citep{jt564} line lists respectively.
They noted that while many of the general features in these spectra were reproduced
by their models, a number of the absorption peaks were shifted. Figure~\ref{fig:Canty}
shows a portion of one of the spectra analysed by  \citet{jt596}. In comparison
we give results of a simple model constructed by us.
The synthetic CoYuTe and BYTe spectra are based on composite
H$_2$O+CH$_4$+NH$_3$ cross-sections (Voigt profile, HWHM 2.0 \cm) which used the
10to10 and POKAZATEL \citep{jt734} line lists for methane and water, respectively.
Relative
molecular abundances are taken from \citet{06SaMaCu.NH3}, which were
also those used by Canty et al. Brown dwarf UGPS 0722 is assumed to have
an effective temperature of 500 K. This is not a full stellar/radiative
transport model so unsurprisingly we do no completely reproduce the observations.
However, a number of observed absorption peaks which were not reproduced by BYTe are indeed present in our model, notably the peaks at 6552 \cm\ and 6588 \cm. We hope that future measurements of high temperature laboratory and astronomical spectra above 6000 \cm\ will further demonstrate the superiority of CoYuTe at shorter wavelengths.

\section{Conclusion}

We present a new line for ammonia, CoYuTe, which replaces our previous line list, BYTe.
Compared to BYTe, CoYuTe covers an increased temperature range, spectral range (extending
to visible wavelengths) and is significantly more accurate. This improved accuracy is achieved
by both starting from an improved potential energy surface and by using empirical energy
levels to improve the line positions for strong transitions.

Use of BYTe allowed the assignment of a number of laboratory ammonia spectra in the near infrared.
However, there remain outstanding issues with ammonia spectra at both near infrared and visible
wavelengths. Hopefully CoYuTe can be used to resolve some these issues and to assign currently unassigned
spectra. Such data would naturally feed through into further improvements in line lists.

\section{Acknowledgement}

This work was supported by the  UK Engineering and Physical Sciences  Research Council (EPSRC)
grant EP/M506448/1 and Servomex Ltd.
We acknowledge support from  the UK Science and Technology Research Council (STFC) No. ST/R000476/1 and COST Action CM1405 MOLIM. A substantial
part of the calculations were performed using high performance computing facilities provided by DiRAC for particle physics, astrophysics and
cosmology and supported by  BIS National E-infrastructure capital grant ST/J005673/1 and STFC grants ST/H008586/1, ST/K00333X/1.
The authors also acknowledge the use of the UCL Legion High Performance Computing Facility (Legion@UCL), and associated support services, in
the completion of this work

\bibliographystyle{mn2e}

\label{lastpage}
\end{document}